%% file: MAIN.tex
\documentclass[a4paper,fleqn]{arxiv}

\usepackage{utils/packages}
\input{utils/commands}

\usepackage{multicol} 
\usepackage[numbers]{natbib}

\definecolor{yellow_star}{rgb}{0.9961, 0.6774, 0.2167}
\definecolor{myblue}{rgb}{0., 0.4, 1}

\def\tsc#1{\csdef{#1}{\textsc{\lowercase{#1}}\xspace}}
\tsc{WGM}
\tsc{QE}

\begin{document}
\let\WriteBookmarks\relax
\def\floatpagepagefraction{1}
\def\textpagefraction{.001}

\shorttitle{Reducing Base Drag on Road Vehicles Using Pulsed Jets Optimized by \HYGO}

\shortauthors{I. Robledo \textit{et al.}}  

\title [mode = title]{Reducing base drag on road vehicles using pulsed jets optimized by hybrid genetic algorithms}

\author[1]{Isaac Robledo}[
    orcid=0009-0005-4426-6219]
\ead{isaac.robledo@alumnos.uc3m.es}
\credit{Methodology, Software, Validation, Formal analysis, Investigation, Data Curation, Visualization, Writing - Original Draft, Writing - Review \& Editing}
\affiliation[1]{organization={Department of Aerospace Engineering, Universidad Carlos III de Madrid, ROR: https://ror.org/03ths8210}, 
            city={Legan\'es},
            postcode={28911}, 
            state={Madrid},
            country={Spain}}
\author[1]{Juan Alfaro}[
    orcid=0009-0006-5744-8772]
\ead{juan.alfaro@alumnos.uc3m.es}
\credit{Methodology, Software, Investigation, Data Curation, Visualization, Writing - Original Draft, Writing - Review \& Editing}
\author[2,1]{Víctor Duro}[
    orcid=0000-0002-9746-7912]
\ead{vdurmer@inta.es}
\credit{Methodology, Investigation, Data Curation, Formal analysis, Visualization, Writing - Review \& Editing}
\author[2,1]{Alberto Solera-Rico}[
    orcid=0000-0003-3883-5989]
\ead{asolric@inta.es}
\credit{Methodology, Investigation, Software, Writing - Review \& Editing}
\author[1]{Rodrigo Castellanos}[
    orcid=0000-0002-7789-5725]
\ead{rcastell@ing.uc3m.es}
\credit{Conceptualization, Methodology, Software, Validation, Formal analysis, Investigation, Data Curation, Writing - Original Draft, Writing - Review \& Editing, Visualization, Supervision, Project administration}
\author[2,1]{Carlos {Sanmiguel Vila}}[
    orcid=0000-0003-0331-2854]
\ead{csanvil@inta.es}
\cormark[1] 
\credit{Conceptualization, Methodology, Software, Validation, Formal analysis, Investigation, Data Curation, Resources, Writing - Original Draft, Writing - Review \& Editing, Supervision, Project administration, Funding acquisition}
\affiliation[2]{organization={Aerial Platforms Department, Spanish National Institute for Aerospace Technology (INTA), ROR: https://ror.org/02m44ak47},
            city={San Mart\'in de la Vega},
            postcode={28330}, 
            state={Madrid},
            country={Spain}}

\cortext[1]{Corresponding author}

\maketitle

\noindent \textsc{Abstract}
\vspace{0.15cm}

Aerodynamic drag on flat-backed vehicles like vans and trucks is dominated by a low-pressure wake, whose control is critical for reducing fuel consumption. This paper presents an experimental study at $Re_W\approx 78,300$ on active flow control using four pulsed jets at the rear edges of a bluff body model. A hybrid genetic algorithm, combining a global search with a local gradient-based optimizer, was used to determine the best-performing jet actuation parameters in an experiment-in-the-loop setup. The cost function was designed to achieve a net energy saving by simultaneously minimizing aerodynamic drag and penalizing the actuation's energy consumption.
The optimization campaign successfully identified a control strategy that yields a drag reduction of approximately 8.8\%. The best-performing control law features a strong, low-frequency actuation from the bottom jet, which targets the main vortex shedding, while the top and lateral jets address higher-frequency, less energetic phenomena. Particle Image Velocimetry analysis reveals a significant upward shift and stabilization of the wake, leading to substantial pressure recovery on the model's lower base. Ultimately, this work demonstrates that a model-free optimization approach can successfully identify non-intuitive, multi-faceted actuation strategies that yield significant and energetically efficient drag reduction.

\vspace{0.4cm}
\noindent \textsc{Keywords}
\vspace{-0.2cm}
\begin{multicols}{2}
\noindent Road vehicle aerodynamics \\ Bluff body \\ Drag reduction \\ Active flow control \\ Pulsed jets \\ Hybrid Genetic Algorithm \\ Machine learning control
\end{multicols}

\newpage
\section{Introduction}\label{s:intro}
Road freight is responsible for approximately 5.2\% of global greenhouse-gas emissions \cite{itf_2021_report}. At highway speeds, roughly 65\% of a vehicle's tractive energy is expended to overcome aerodynamic drag \cite{mccallen2004_aerodynamic_pollution_contribution}. For flat-backed vehicles such as vans and lorries, this drag is dominated by a low-pressure region on the base, which results from flow separation at the sharp trailing edges. Consequently, mitigating this base-pressure deficit through wake control provides a direct and effective means to reduce fuel consumption and the associated emissions.

The wakes of simplified ground vehicles like the Ahmed body have been the subject of extensive research, revealing a highly complex, three-dimensional flow field that is acutely sensitive to geometric variations \cite{Grandemange2013WakeBluntBody}. The near-wake is typically characterized by two large recirculation regions originating from the shear layers that separate from the upper and lower base edges \cite{cerutti2020VanCafiero}. In the lateral direction, the wake often exhibits a global symmetry breaking, with modal switching between quasi-symmetric and anti-symmetric states. This low-frequency dynamic, linked to the shedding of large-scale vortex structures, produces broadband pressure fluctuations and a depressed time-averaged base pressure. Foundational experimental and numerical studies have thoroughly mapped the bi-modal character of these wakes and their sensitivity to geometry and operating conditions \cite{dalla2019simulationsBimodalFlatBackBluff,fan2020experimentalBiStableFlatBackAhmed,khan2024equilibriumFluxesFlatBackBluffBody, Grandemange2013BistabilityBackAR}, providing a baseline understanding that informs all modern control strategies.

Modifying these wake dynamics offers a promising route for drag reduction. Passive control devices have been widely explored, including vertical flaps to attenuate lateral vortex structures \cite{Beaudoin2008flaps}, base cavities to improve performance in crosswinds \cite{urquhart2021drag}, splitter plates \cite{gillieron2010aerodynamic}, boat tails \cite{lanser1991aerodynamic, khalighi2001experimental}, and fences \cite{modi1995drag}. In contrast, active flow control provides greater authority due to its adaptive capabilities to changing conditions \cite{yu2021recentAdvances} and has been investigated extensively through momentum injection at the trailing edges \cite{mcnally2015drag}.
In particular, strategies involving steady momentum injection have proven effective, with \citet{roumeas2009drag} achieving a 17\% drag reduction on an Ahmed body via suction prior to separation, and \citet{aubrun2011separation} obtaining up to 14\% drag reduction using an array of steady blowing microjets. A widespread and often more efficient alternative is unsteady forcing, through pulsed or synthetic jets. These actuators can promote flow reattachment, as explored by \citet{glezer2005aspects}, or directly target wake instabilities. Examples of the latter include parametric studies to optimize actuation on the slanted surfaces of an Ahmed body \cite{park2013aerodynamic, joseph2013flow}, and targeted base pressure recuperation using high-frequency forcing \cite{oxlade2015high}. Some approaches have even combined passive devices with active suction and blowing to maximize performance \cite{seifert2015lab}.

More recently, learning-based, model-free approaches have gained traction, offering powerful tools to navigate the vast and complex parameter space of active flow control laws directly on experimental setups. This paradigm bypasses the need for explicit, often intractable, low-dimensional models of turbulent wakes, allowing algorithms to discover effective, sometimes non-intuitive, control strategies through interaction with the physical system \cite{li2017GP_FlatBackAhmed, Zhang2023AIAhmed,zhang2025autonomousrealtimecontrolturbulent}. Several successful applications have demonstrated the potential of these methods on benchmark automotive geometries, primarily the Ahmed body. For instance, \citet{li2017GP_FlatBackAhmed} employed linear genetic programming (LGP) to optimize pulsed jets combined with Coanda surfaces on a square-back Ahmed body, achieving a remarkable 22\% drag reduction. Their model-free approach identified optimal multi-frequency forcing laws. \citet{Zhang2023AIAhmed} utilized an Ant Colony Optimization (ACO) algorithm to control distributed steady microjet arrays on a low-drag Ahmed body (slant angle $\varphi=35^{\circ}$), reducing drag by 18\% while explicitly incorporating control power input into the cost function to seek efficient solutions. \citet{deng2023sensitivity} applied an Explorative Downhill Simplex Method (EDSM) to optimize independently operated pulsed microjets on a square-back Ahmed body, achieving 13\% drag reduction. Their work particularly focused on sensitivity analysis for a large number of control parameters (up to 12) and demonstrated the potential for significantly enhanced control efficiency (up to 78\%) with only a minor sacrifice in drag reduction. Demonstrating a fully autonomous approach, \citet{zhang2025autonomousrealtimecontrolturbulent} developed REACT, a Reinforcement Learning framework deployed experimentally on a square-back Ahmed body using servo-actuated flaps. Learning directly in the wind tunnel from onboard pressure sensors, REACT achieved a 3.64\% drag reduction with net energy savings, discovering that dynamically suppressing wake instabilities was the optimal strategy. Notably, their physics-informed training allowed the agent to generalize across different flow speeds without retraining. Collectively, these successes underscore the competitive performance and adaptability of machine learning control for manipulating complex wake flows on standard automotive shapes

Moving beyond the standard Ahmed body benchmark, active flow control strategies have also been applied to more realistic vehicle shapes, notably the van model geometry utilized in \cite{amico2022geneticFurgo,amico2022deep,amico2024flow}. In a first study \citep{amico2022geneticFurgo}, a Genetic Algorithm (GA) was used to optimize multi-frequency pulsed jets in a square-back van (represented as a sum of two sine waves), achieving an 11.2\% drag reduction. Notably, their initial cost function aimed purely at maximizing drag reduction without explicitly penalizing actuation energy. The follow-up work \cite{amico2024flow} employed Deep Reinforcement Learning (DRL), exploring different state representations (using base pressure) and reward definitions, including cases explicitly designed to consider the energy budget alongside drag reduction. Their subsequent flow topology analysis revealed distinct wake modifications depending on whether the DRL agent prioritized maximal drag reduction or energy efficiency. The efficiency and adaptability of DRL have made it one of the most attractive closed-loop control methodologies not only for bluff-body flows, but more broadly within fluid mechanics. Representative examples include closed-loop attenuation of a cylinder wake using plasma actuators \citep{Zong2024airfoilplasmaclosedloop}, airfoil separation control \citep{Zong2024airfoilplasmaclosedloop}, and highly complex configurations such as the maximization of shear-layer mixing in a supersonic backward-facing step \citep{Zong2025closedloopsupersonic}.

Despite the demonstrated promise, the efficacy and practical applicability of these learning-based strategies can be constrained by several factors. Firstly, the reliance on a single optimization algorithm (be it GA, LGP, ACO, EDSM, or DRL) within many individual studies risks incomplete exploration of the potentially vast and multimodal control parameter space, possibly leading to convergence towards locally optimal, rather than globally optimal, solutions. Secondly, a critical consideration, often omitted in earlier works but highlighted in more recent ones, is the actuation cost. Neglecting the energy expenditure of the control system within the optimization loop can yield strategies that, while achieving substantial drag reduction, are energetically inefficient or entirely impractical for real-world applications. While several studies have begun incorporating energy considerations \citep[e.g.][]{Zhang2023AIAhmed,deng2023sensitivity,amico2024flow}, it remains a crucial factor for practical viability. Finally, many experimental implementations, particularly those involving hardware-in-the-loop optimization, do not formally account for the propagation of measurement uncertainty. The inherent noise in experimental data, if not properly managed, can be amplified, introducing significant bias into the learning process and potentially guiding the optimization towards spurious or non-robust outcomes.

To address these limitations, the present study introduces a novel hybrid genetic algorithm that synergistically combines the global exploratory power of a genetic algorithm with the local exploitation capabilities of the Downhill Simplex Method, a local search algorithm. We apply this algorithm experimentally to a simplified van model in a wind tunnel as in \citet{amico2022geneticFurgo,amico2022deep,amico2024flow}, optimizing the control parameters of a pulsed-jet system while simultaneously characterizing the resulting wake modifications. Crucially, the objective function incorporates a penalty term for the momentum injected by the actuators, ensuring that the optimization converges towards energetically feasible solutions. Furthermore, to guarantee the robustness of the experimental evaluation, each candidate solution is subjected to repeated measurements, and a statistical uncertainty threshold is employed to discard unreliable data, thereby providing a high degree of confidence in the final performance metrics.

The remainder of this paper is organized as follows. Section~\ref{s:Methodology} details the experimental setup, including the wind tunnel facility, the bluff body model, and the actuation and measurement systems. Section~\ref{s:machine_learning_control} presents the hybrid genetic algorithm, defining the control parameterization, the cost function, and the methodology for handling experimental uncertainty. The results of the optimization are presented and discussed in Section~\ref{s:results}, where the best-performing control strategies is analyzed in detail, and the corresponding flow physics are examined in Section~\ref{s:piv_analysis}. Finally, Section~\ref{s:conclusions} summarizes the key findings of the study and offers concluding remarks.

\begin{figure*}[ht]
    \centering
    \includegraphics[width=1\linewidth]{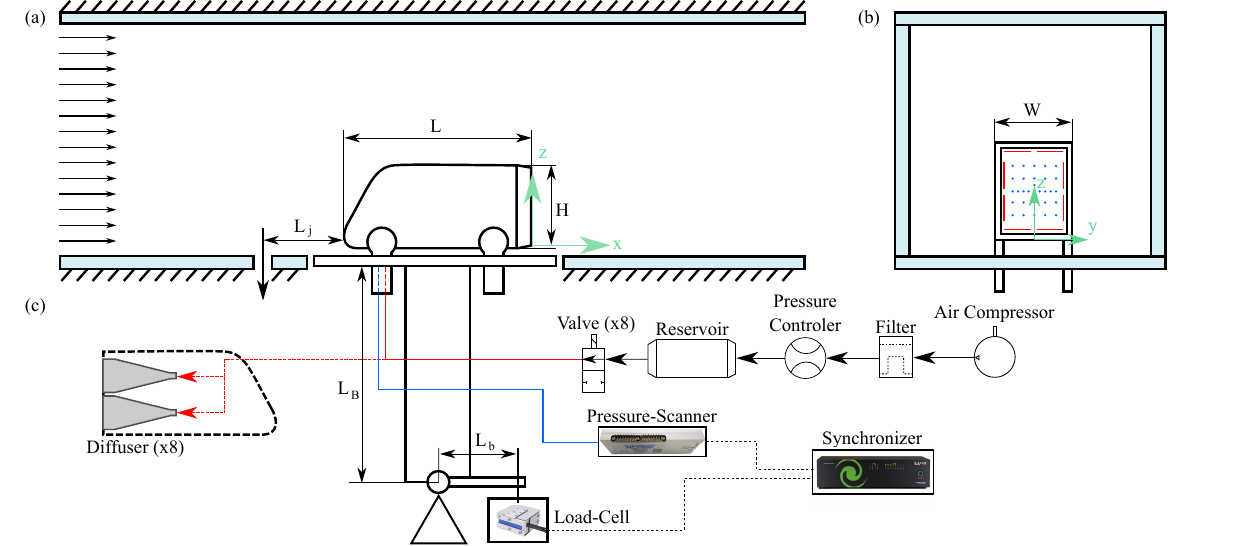}
    \caption[]{Schematic of the experimental setup. (a) Side view of the wind tunnel test section showing the bluff body model. (b) Rear view of the model's base, indicating the locations of the 31 static pressure taps (blue circles \tikz\draw[fill=myblue, draw=myblue] (0,0) circle (3.pt);) and the four pairs of pulsed jet actuators \lcap{-}{red}. (c) Diagram of the auxiliary systems, including the pneumatic circuit for actuation, the force balance and load cell for drag measurements, and the pressure data acquisition system.}
    \label{fig:exp_setup}
\end{figure*}
\section{Experimental setup and Methodology} \label{s:Methodology}
This section describes the experimental setup and measurement techniques employed in this study. The pneumatic actuation system, the force, and pressure acquisition systems are presented together with the Particle Image Velocimetry flow field setup.

\subsection{Experimental Facility and Model Setup} \label{ss:exp_facility}
The experiments were conducted in an Eiffel-type wind tunnel. The facility has a test section with a square cross-section of $0.4 \times 0.4$~m$^2$ and a length of $2$~m. All measurements were performed at a free-stream velocity of $U_\infty=12$~m/s, with a corresponding turbulence intensity below 0.3\%.

The bluff body is a scaled version of the van model used in \citet{cerutti2020VanCafiero} and \citet{amico2022geneticFurgo,amico2022deep,amico2024flow}, with a height of $H=0.12$ m, a width of $W=0.102$ m, and a length of $L=0.247$ m. The clearance between the model and the wind tunnel floor is equal to $h=0.012$ m, leaving a normalized value $h/H=0.1$. This configuration results in a wind tunnel blockage ratio of 7.65\% and a Reynolds number based on width of $Re_W \approx 78,300$. It is noted that, even though the external geometry of the model is identical to the referenced study, the interface with the load transducer and the actuation systems is not the same (see \S\ref{ss:actuation}). As depicted in \autoref{fig:exp_setup}, the model is positioned on a free-floating platform, maintaining a lateral clearance of approximately $1$ mm with respect to the wind tunnel floor. It is rigidly mounted to the platform through rectangular extensions from the wheels, designed so that the wheel base would be flush with the floor. This mounting strategy provides a non-intrusive passage for the tubing of the actuation and instrumentation systems, thereby precluding the need for internal support structures and the associated flow corrections.

\subsection{Actuation System}\label{ss:actuation}
The actuation system comprises four pairs of parallel slots positioned on the four edges of the model's base, as indicated by the red lines in \autoref{fig:exp_setup}, (b). This arrangement is analogous to actuator configurations used in recent bluff-body flow-control studies, such as the single-slot jets employed by \citet{amico2022geneticFurgo}. Each rectangular slot features a cross-section of $39.4 \times 1$~mm$^2$. The use of slot pairs, separated by $5$~mm on each side of the model, is employed to improve the homogeneity of the jet outflow. 

Actuation is driven by eight Matrix MX 821.104C2KK solenoid valves, with each pair of valves electrically connected to a single controller channel. Each valve is triggered by a $24$~V periodic square signal, characterized by a carrier frequency $f$ and a duty cycle $DC$ (or ratio between pulse width and signal period). These valves feature a maximum operating frequency of $500$~Hz, and a response time of less than $1.3 \pm 0.3$~ms. The pneumatic supply is common to all actuators. An Alicat Scientific\textsuperscript{\texttrademark} M-500SLPM mass flow controller sets the system's pressure at a constant $P_j=5$~bar, while monitoring both mass flow rate, absolute pressure, temperature, and other metrics. Downstream of the controller, a secondary reservoir dampens pressure oscillations before the flow is split into eight individual tubes, one for each valve. Finally, each valve feeds a dedicated tube connected to a diffuser, which is rigidly integrated into the bluff body's base to discharge the pulsed jets into the wake. This configuration yields a jet exit velocity ratio of $V_{jet}/U_\infty \sim 1$.

To ensure a well-defined and repeatable inflow condition, a boundary layer suction system was integrated into the wind tunnel floor. This system, comprised by three dual-axial fans, prevents the thick, naturally-developing tunnel floor boundary layer from impinging on the model, which could otherwise lead to unrealistic flow phenomena. A transverse slot with a width of $7$~mm and a length of $215$~mm, is located $L_j = 98.5$~mm upstream of the bluff body's leading edge. This device effectively removes the incoming boundary layer, ensuring that the flow approaching the model is representative of on-road conditions without introducing significant disturbances to the freestream profile.

\subsection{Force and base Pressure Measurements}\label{ss:force_flow_measurements}

Aerodynamic drag force is measured via a custom lever system acting on a Fibos FA702 three-axis load cell. The body-platform assembly, described in Section~\ref{ss:exp_facility}, is mounted on a primary cantilever beam of length $L_B = 0.82$~m. This beam pivots on a high-precision bearing, a mechanism that translates the streamwise aerodynamic force exerted on the model into a vertical load on a secondary lever arm of length $L_b = 85$~mm. The resulting force is transmitted via a vertical threaded rod to the load cell. In the primary measurement axis, the load cell has a maximum capacity of $5$~N and sensitivity $1\;mV/V$ with an excitation of $12\;V$. Data from the load cell is acquired using a Viking VK702NH data acquisition system, which provides a $10$~V excitation voltage and utilizes a 24-bit ADC with an input range of $100$~mV, sampling at $800$~Hz. Finally, solid blockage \cite{barlow1999low} and wake blockage correction using Maskell’s methodology \cite{Gould1969WakeBlockage} is applied to the measured drag coefficients.

Time-resolved base pressure distribution was acquired using a 32-channel Scanivalve MPS4232 pressure scanner, with synchronous sampling capability across all channels, and a maximum sampling frequency of $1000$~Hz. The Scanivalve measures pressure differentials with respect a reference value, which in the case of this study was selected as $P_\infty$ such that the obtained measurements are directly the pressure differential $\Delta P = P-P_\infty$.
31 pressure ports are distributed across the model's base, with the pattern shown as blue dots in \autoref{fig:exp_setup} (b). This arrangement consists of a central $5\times 5$ square grid with a horizontal spacing of $\Delta x=15$~mm and a vertical spacing of $\Delta y=20$~mm. The spatial resolution was enhanced along the horizontal and vertical centerlines by adding intermediate pressure taps between the main grid points.
\begin{figure}
    \centering
    \includegraphics[width=0.95\linewidth]{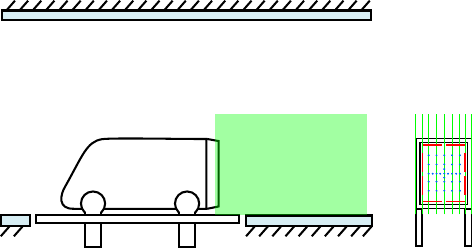}
    \caption[]{Schematic of the PIV planes, illustrating the field of view $-0.03\leq|x|/W\leq\ 1.82$ and $-0.112\leq|z|/W\leq 1.433$ and the lateral positioning. A total of nine planes were acquired symmetrically distributed about the base's centerline, which include two planes aligned with the model's side walls ($|y|/W=0.5$), two aligned with the actuators' ($|y|/W=0.43$), four positioned in the vertical lines defined by the pressure probes ($|y|/W=0.29$ and $|y|/W=0.15$), and one plane in the centerline ($|y|/W=0$).}
    \label{fig:exp_setup_piv}
\end{figure}

A dedicated synchronizer (iLA 5150) was employed to trigger the simultaneous acquisition of data from both the load cell and the pressure scanner at a final sampling rate of $800$~Hz.

\subsection{Velocity Measurements}\label{ss:PIV}

Two-component Particle Image Velocimetry (PIV) was used to measure the streamwise and wall-normal velocity fields in several vertically-aligned planes, as depicted in \autoref{fig:exp_setup_piv}. The flow was seeded with Di-Ethyl-Hexyl-Sebacate (DEHS) particles with a nominal diameter of approximately 1~$\mu$m. Illumination was provided by a dual-cavity Nd:Yag Quantel Evergreen laser (200 mJ/pulse at 15 Hz), with the light sheet formed by a set of cylindrical and spherical lenses. An iLA5150 sCMOS camera, fitted with a 50~mm lens set to a focal ratio of $f_\#=11$, captured the particle images. The resulting field of view covered a region of $1.85\times 1.545$~mm$^2$ ($-0.03\leq x /W\leq\ 1.82$ and $-0.112\leq z/W\leq 1.433$) with a spatial resolution of approximately 13.3 pixels/mm.

Measurements were acquired in nine distinct planes, as depicted in \autoref{fig:exp_setup_piv}, distributed symmetrically about the model's centerline. Five of these planes were aligned with the central columns of pressure taps, separated by $\Delta x=15$~mm. Two additional planes were positioned to capture the flow directly downstream of the vertical actuator slots, and the final two were aligned with the model's lateral walls.

For each plane, flow statistics were computed from an ensemble of 500 image pairs acquired at a sampling frequency of 15~Hz. The raw images were processed using a multi-pass, multi-grid cross-correlation algorithm with window deformation \cite{SORIA1996piv, scarano2001piv}. The interrogation concluded with a final window size of $32 \times 32$~pixels$^2$ and a $75$\% overlap, yielding a vector spacing of $2.4$~vectors/mm. The time separation between laser pulses was set to 74~$\mu$s to accommodate the large velocity gradients and low seeding density near the unseeded jets. This resulted in typical particle displacements of $12$~pixels, with an estimated uncertainty of 0.1~pixels for the displacement field \cite{Raffel2018piv}.

The Cartesian coordinate system $(x, y, z)$ is defined such that the origin lies at the midpoint of the lower edge of the bluff body's base, aligned with the symmetry plane (see \autoref{fig:exp_setup}). Here, $x$, $y$, and $z$ represent the streamwise, spanwise, and wall-normal directions, respectively. Since only vertical PIV planes are acquired in the study, only the stream-wise $U$ and wall-normal velocities $V$ are labelled, with the corresponding magnitude $||U||$. The metrics represented through the paper are normalized with respect to $U_\infty$. The time-averaged statistics are labelled $\overline{\bullet}$, while the fluctuating measurements are $\Phi' = \overline{|\Phi - \overline{\Phi}|}$. 

\begin{table}
\centering
  \caption{Optimization hyperparameters for HyGO}
  \label{tab:optimization_params}
  \begin{tabular}{ccl}
    \toprule
    Name & Value & Description\\
    \midrule
    \midrule
     $N_b^{f}$ & 7 & Bit number for $f$\\
     $N_b^{DC}$ & 5 & Bit number for $DC$\\
     $N_b^{\phi}$ & 4 & Bit number for $\phi$\\
     $N_g$ & 10 & Total number of generations\\
     $N_{init}$ & 100 & Initialization size\\
     $N_{explor}^2$ & 75 & Gen 2 Exploration size\\
     $N_{explor}^{3-10}$ & 50 & Gens 3 - 10 Exploration size\\
     $N_T$ & 7 & Tournament size\\
     $p_T$ & 1 & Tournament selection probability \\
     $P_c$ & 0.55& Crossover probability\\
     $P_m$ & 0.45& Mutation probability\\
     $p_m$ & 0.05& Mutation Bit flip Probability\\
     $P_r$ & 0   & Elitism probability\\
     $N_{exploit}$ & 20\% $N_{explor}$ & Exploitation population Sizes\\
     $N_s$ & 8 & Simplex Size\\
    \bottomrule
\end{tabular}
\end{table}

\section{Machine learning control}\label{s:machine_learning_control}

Building on the facility and diagnostics detailed in Section \ref{s:Methodology}, we formulate drag reduction as an open-loop, model-free optimization task over the space of actuation policies \cite{BruntonNoack2015review,duriez2017book}. Given the noisy, nonconvex, and potentially discontinuous nature of the objective landscape, we employ a genetically inspired hybrid genetic algorithm that combines global exploration with local refinement, thereby enhancing convergence while exploiting local minima \cite{Robledo2025Hygo}. This section introduces the problem formulation, details the scaling used to construct the cost function, and outlines the algorithmic framework and training process.

\subsection{Formulation of the optimization problem}

We formulate drag reduction as an open-loop, model-free optimization over a parametrized family of actuation schedules. Let $\mathbf{b}(t)\in\mathbb{R}^{N_b}$ be the actuation vector on time $t$ and let $\mathbf{K}(t;\theta):[0,T]\to\mathbb{R}^{N_b}$ denote a control law specified by a parameter vector $\theta\in\mathbb{R}^{N_p}$. Fixed experimental and flow conditions are collected in $\Theta$. The goal is to select $\theta$ so that the induced schedule $\mathbf{b}(t;\theta)=\mathbf{K}(t;\theta)$ minimizes a scalar cost $J$ that aggregates performance and penalties (the construction of $J$ is detailed in Subsection \ref{ss:cost_function}).
\begin{equation}
\mathbf{K}^* = \arg \min_{\mathbf{K} \in \mathcal{K}} J(\mathbf{K}(t;\theta);\Theta)
\end{equation}
where $\mathcal{K}$ is the admissible space of actuation commands, bounded by the limits of the parameters $\theta$.

The actuation system consists of four pairs of slotted jets located at the rear of the bluff body, with assigned control laws $b_1$ through $b_4$ in clockwise order starting from the top actuator (see \autoref{fig:exp_setup}). To reduce dimensionality, the lateral symmetry of the problem is exploited: the side actuators ($b_2,\;b_4$) are constrained to share the same frequency and duty cycle (denoted $f_2,\;DC_2$), while retaining a relative phase shift $\phi$. The top ($b_1$) and bottom ($b_3$) actuators remain independent, each with their own $f$ and $DC$. In total, the optimization problem involves seven control parameters:
\begin{equation}\label{eq:parameters}
    \theta = \left[f_1,DC_1,f_2,DC_2,f_3,DC_3,\phi\right]
\end{equation}
We introduce a phase shift only between the lateral actuators since relative phase is ill-defined, and thus not a meaningful control variable, when the signals have different frequencies \cite{Wu2018jet,minelli2020lgac}. Flat-back bluff bodies, however, exhibit a strong lateral vortex-shedding mode with a well-defined dominant frequency \cite{evrard2016fluidSymmetryModes,dalla2019simulationsBimodalFlatBackBluff,fan2020experimentalBiStableFlatBackAhmed,khan2024equilibriumFluxesFlatBackBluffBody}. In particular, our model ($H/W\approx1.17$) is expected to present both lateral and vertical shedding instabilities \citep{Grandemange2013BistabilityBackAR} (but dominated by the vertical one). By constraining the lateral actuators to share the same frequency $f_2$ and duty cycle $DC_2$ while varying their relative phase $\phi$ we specifically target the lateral, symmetric instability mode, which has been shown \cite{KHAN2022ControlBistable} to contribute significantly to drag when present, and whose partial attenuation, rather than complete suppression, was demonstrated to yield effective drag reduction.

It is also noted that, in the present parametrization, the duty cycle $DC_2$ simultaneously controls two actuators, whereas $DC_1$ and $DC_3$ each affect only a single slot. As a result, variations in $DC_2$ have a larger impact on the total injected mass flow and on the global cost function, which implies a higher optimization sensitivity to this parameter and may inherently bias the search toward reducing $DC_2$ relative to the other duty cycles.
\begin{table*}
\centering
  \caption{Performance metrics and parameters for the reference cases of no actuation and steady-jet, together with the best performing and minimal $J_a$ individuals.}
  \setlength{\tabcolsep}{11pt} 
  \label{tab:best_results}
  \begin{tabular}{lcccccccccc}
    \toprule
    Name & $J$ & $J_a$ & $J_b$ & $\tilde{f}_1$ & $\tilde{f}_2$ & $\tilde{f}_3$ & $DC_1$ & $DC_2$ & $DC_3$ & $\phi$\\
    \midrule
    \midrule
    No Actuation & $1$ & $1$ & $0$ & $-$ & $-$ & $-$ & $-$ & $0$ & $0$ & $0$\\
    Steady-jet & $1.193$ & $1.011$ & $1$ & $-$ & $-$ & $-$ & $1$ & $1$ & $1$ & $-$\\
    Best $J$ & $1$ & $0.912$ & $0.486$ & $0.848$ & $0.855$ & $0.119$ & $0.427$ & $0.282$ & $0.573$ & $24$\\
    Best $J_a$ & $1.02$ & $0.89$ & $0.716$ & $0.614$ & $0.761$ & $0.58$ & $0.686$ & $0.524$ & $0.669$ & $120$\\
    \bottomrule
\end{tabular}
\end{table*}

Frequencies are reported in non-dimensional form using the Strouhal number
\begin{equation}
    St \;=\; \tilde{f} \;=\; \frac{f\,W}{U_\infty}.
\end{equation}

The actuation commands are defined as binary (on/off) signals obtained by thresholding biased sinusoids:
\begin{equation}
    \begin{split}
        b_i(t) &=h\big(\sin(\omega_i t)+\kappa(DC_i)\big),\quad\quad\quad i=1,2,3\\
        b_4(t) &= h\big(\sin(\omega_2 t+\phi)+\kappa(DC_2)\big),
    \end{split}
\end{equation}
where $h(\cdot)$ denotes the Heaviside step function ($h(x)=1$ if $x>0$, and $0$ otherwise), 
$\omega_i = 2\pi f_i = 2\pi \tilde{f}_i\,U_\infty/W$, and $\tilde{f}_i$ is the non-dimensional frequency (Strouhal number) of actuator $i$. 
The bias $\kappa(DC)$ maps a desired duty cycle $DC\in(0,1)$ to the sinusoidal offset, ensuring that the fraction of time with $b_i(t)=1$ over a period equals $DC$ exactly.

The admissible parameter space is $\tilde{f}_i\in[0.085,\,0.935]$, $DC_i\in[0.25,\,0.75]$, $\phi\in[0,\,\pi]$. Each parameter's range is uniformly discretized into $2^{N_b}$ values, where the number of bits, $N_b$, is specified for each parameter in \autoref{tab:optimization_params}. These boundaries are set to concentrate the search around the expected lateral shedding frequency of flat-back bluff bodies, $\tilde{f}_{\text{shed}}\approx 0.11\;\text{–}\;0.20$ \cite{li2017GP_FlatBackAhmed,fan2020experimentalBiStableFlatBackAhmed}.

\subsection{Cost function}\label{ss:cost_function}
The objective function must primarily reflect drag reduction. However, directly minimizing a drag-related quantity can bias the optimizer toward control strategies with excessive mass or momentum injection (i.e., high duty cycles), artificially improving performance metrics without necessarily targeting the underlying flow mechanisms responsible for genuine drag mitigation. To prevent this, the cost function includes an explicit penalty on the injected mass flow rate (as discussed by \citet{castellanos2023genetically} for heat transfer), discouraging unphysical solutions and ensuring that the reported improvements result from effective flow control rather than disproportionately strong actuation. Both objectives are combined into a single, dimensionless cost function to allow consistent comparison across all cases:
\begin{equation}
    J(\theta) \;=\; J_a \;+\; \gamma\cdot J_{b}
    \;=\; \frac{D}{D_0} \;+\; \gamma\cdot\frac{\dot m}{\dot m_{\mathrm{SJ}}}
\end{equation}
where $D$ and $\dot m$ are the episode-averaged (whose duration is defined in \ref{s:IndividualEvaluation}) drag and mass-flow rate, $D_0$ is the mean drag with no actuation, and $\dot m_{\mathrm{SJ}}$ is the mean mass flow when all actuators are continuously on (hereafter, the steady-jet reference). 

The dimensionless drag term, $J_a = D/D_0$, quantifies the effectiveness of each candidate actuation (lower values indicate better performance) and depends on all control parameters. The penalization term, $J_b = \dot{m} / \dot{m}_{\mathrm{SJ}}$, represents the injected mass flow relative to the steady-jet reference. Under binary actuation with fixed supply conditions, $J_b$ depends solely on the duty cycles, being independent of frequency and phase. With this normalization, the no-actuation baseline yields $J = 1$. An in-depth description of the evaluation procedure for each individual is provided in \autoref{s:IndividualEvaluation}.
The weight $\gamma \ge 0$ defines the relative contribution of the mass-injection term $J_b$ to the global cost function $J$. In the present formulation, $\gamma$ is not only introduced as a subjective trade-off parameter between aerodynamic performance and actuation cost, but also to decouple genuine wake-induced drag reduction from artificial improvements derived from excessive mass injection. When $\gamma = 0$, the optimizer systematically favors solutions with large $J_b$, leading to a reduction of $J_a$ that is not necessarily associated with beneficial wake modification. Conversely, excessively large values of $\gamma$ drive the optimization toward vanishing actuation, suppressing physically meaningful control strategies. The role of $\gamma$ is therefore set to compensate for this mass-injection effect and to ensure that the ranking of individuals is primarily governed by their ability to modify the wake and recover base pressure.
Following \citet{castellanos2023genetically}, we initialized the optimization using 100 Latin Hypercube samples (that are later taken as the first generation for the optimization) to probe the parameter space and quantify the correlation between $J_a$ and $J_b$. For $\gamma = 0$, these samples exhibit a strong linear dependence between mass injection and apparent drag reduction, indicating that higher blowing levels trivially improve $J_a$ through mass injection. The value $\gamma = 0.182$ was obtained by linear regression so as to flatten this trend, effectively neutralizing the direct contribution of mass injection to the cost function. Although the initial LHS sampling is necessarily sparse in a seven-dimensional experimental space, the robustness of this choice is validated by the convergence of the optimization toward non-trivial solutions. All subsequent results, therefore, employ this fixed value of $\gamma$.

\subsection{Hybrid genetic algorithm optimization}\label{ss:hygo_intro}
Given the complex optimization landscape, featuring nonlinear couplings and collinearity, possible non-smoothness, and a moderately high dimensionality, we employ a hybrid genetic algorithm. The optimizer, labelled HyGO and introduced by \citet{Robledo2025Hygo} includes both GA and GP functionalities but, due to the open-loop and parametric nature of the target problem, the hybridized GA was employed. Furthermore, HyGO accelerates convergence by appending a local-refinement step to each generation while retaining the global search capability of standard GAs.

An individual encodes the parameter vector in Equation~\eqref{eq:parameters}. As stated in Section \ref{ss:cost_function}, the search is initialized via Latin Hypercube Sampling with $N_{init}=100$ individuals. Each subsequent generation proceeds in two stages. First, an \emph{explorative} stage (generation 1 uses the LHS set as its exploration pool) applies standard GA operators to produce $N_{explor}$ offspring, which are then evaluated and ranked. Second, an \emph{exploitative} stage generates $N_{exploit}$ additional candidates by locally refining selected high-ranked individuals using the Downhill Simplex method (Nelder–Mead), chosen for its gradient-free and fast-convergent behavior. The number of generations $N_g$, tournament size for exploration, probabilities of genetic operators, and other hyperparameters are reported in \autoref{tab:optimization_params} and were selected based on prior experience and related studies \cite{Yu2021GA_drag_slot_jets, cornejo2021gMLC, castellanos2023genetically}. We employ a decreasing exploration population across generations, which front-loads exploration to sample multiple local minima early, and progressively shifts budget toward local refinement.

\begin{figure*}
    \centering
    \includegraphics[width=1\linewidth]{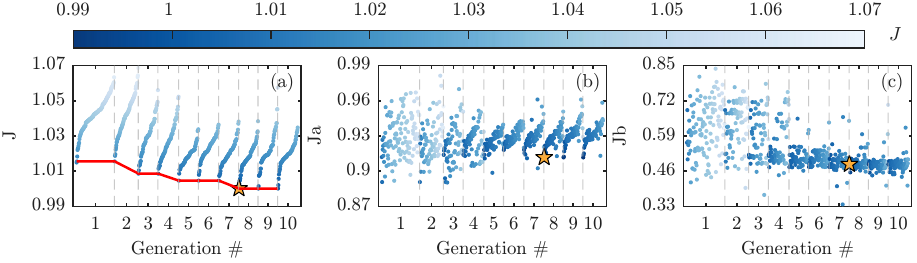}
    \caption[]{Evolution of the optimization process across generations, showing: (a) the total cost, $J$; (b) the drag cost, $J_a$; and (c) the penalization, $J_b$. Within each generation, individual solutions are represented by circular markers, which are sorted and coloured according to their total cost $J$ in ascending order from left to right. The yellow star \tikz\node[star, fill=yellow_star, draw=black, star point ratio=2.5, inner sep=1.5pt] {}; denotes the best-performing individual in the optimization, $\min(J)$, placed within the generation it appears.}
    \label{fig:J_evolution}
\end{figure*}
\begin{figure}
    \centering
    \includegraphics[width=0.95\linewidth]{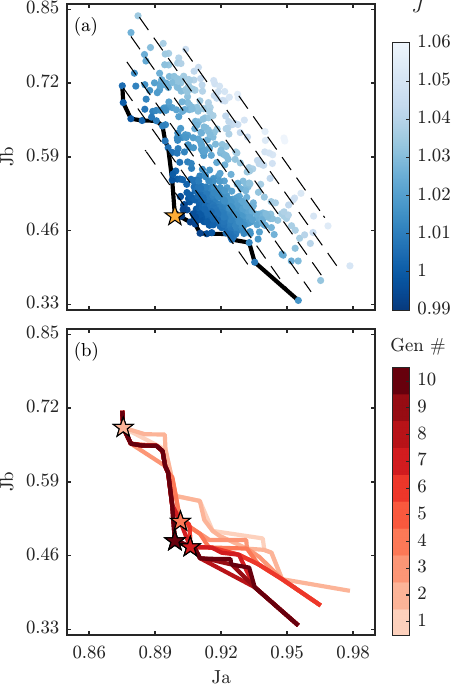}
    \caption[]{Analysis of the optimization, illustrating the relationship between aerodynamic performance ($J_a$) and actuation cost ($J_b$). (a) Distribution of all evaluated individuals in the objective space, plotting drag cost, $J_a$, against penalization, $J_b$. The data points are colored by their total cost, $J$, and the dashed lines \lcap{--}{black} represent iso-contours of constant $J$. The solid black line \lcap{-}{black} indicates the non-dominated, or Pareto, front, with the yellow star \tikz\node[star, fill=yellow_star, draw=black, star point ratio=2.5, inner sep=1.5pt] {}; marking the location of the overall fittest individual. (b) Evolution of the Pareto front across successive generations, indicated by color. The stars (\tikz\node[star, fill=white, draw=black, star point ratio=2.5, inner sep=1.5pt] {};) denote the best-performing individual discovered up to that respective generation.}
    \label{fig:Pareto}
\end{figure}
\begin{figure*}
    \centering
    \includegraphics[width=0.85\linewidth]{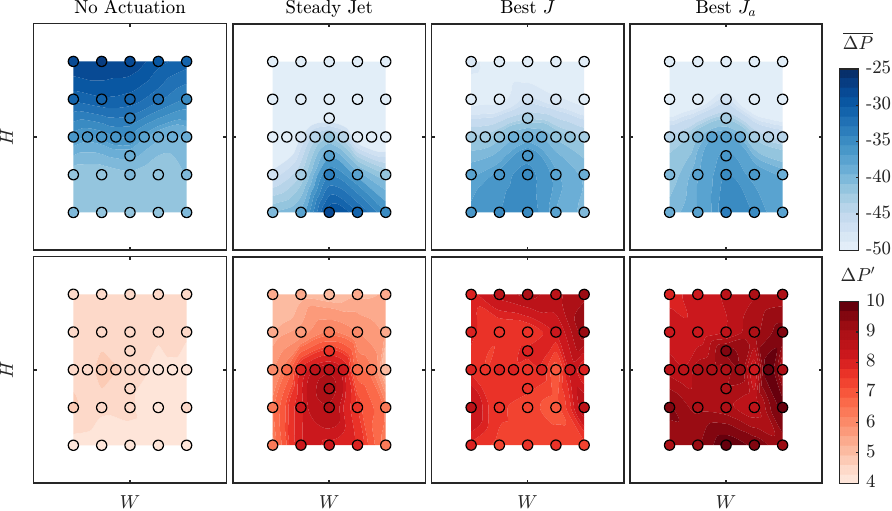}
    \caption{Time-averaged ($\overline{p}$) and fluctuating ($p'$) base pressure contours for four characteristic cases. From left to right, the columns correspond to: (i) the non-actuated baseline flow; (ii) continuous steady-jet actuation; (iii) the best-performing individual achieving the minimum total cost, $\min(J)$; and (iv) the individual achieving the maximum drag reduction, $\min(J_a)$. Each image limits represent entire base surface.}
    \label{fig:Averages}
\end{figure*}

\section{Results}\label{s:results}
\subsection{Optimization results}
A total of $688$ individuals were evaluated over $10$ generations. Thanks to the uncertainty-screening protocol described in the Appendix~\ref{s:IndividualEvaluation}, no outliers were retained in the analysis. The hybrid optimizer exhibited fast convergence: it explored the parameter space broadly in the early generations and then converged rapidly in all parameters except $\phi$. Convergence is further supported by the plateau of the best objective in the final three generations (\autoref{fig:J_evolution}(a)), where no further improvement is observed.

A clear optimum region emerges for the three actuation frequencies: $\tilde{f}\approx0.85$ for the top and side slots ($\tilde{f}_1,\tilde{f}_2$) and $\tilde{f}_3\approx0.12$ for the bottom slot, the latter lying close to the natural shedding frequency of the wake. Some scatter remains in $\tilde{f}_1$, with high-performing individuals sparsely distributed across its domain. Furthermore, the optimizer tends toward intermediate blowing levels for the top and bottom diffusers while minimizing the side blowing. Since $DC_2$ regulates two slots (whereas $DC_1$ and $DC_3$ each regulate one), the objective $J$ may implicitly favor reducing $DC_2$ to limit total mass flow, leaving drag reduction to the other actuators. This remains a hypothesis; targeted sensitivity tests would be required to confirm it. Consistently with this reduced lateral authority, the best-performing solutions exhibit a large scatter in the lateral phase $\phi$, with both $\phi \approx 24^\circ$ and $\phi \approx 120^\circ$ yielding comparable performance in the best $J$ and $J_a$ solutions, indicating that the objective function is only weakly sensitive to the side-slot phase in the vicinity of the found minimum.

Interestingly, $J_a$ attains low values from the beginning, so HyGO prioritizes maintaining $J_a$ at a “good-enough” level while reducing the momentum input, thereby targeting genuine drag reduction mechanisms, an evolution not immediately apparent in \autoref{fig:J_evolution}(b). The Pareto analysis in \autoref{fig:Pareto}(b) corroborates this behavior: the best early individuals lie near the minimum of $J_a$, and subsequent generations progress downward along the right end of the front, achieving lower $J_b$ at nearly constant $J_a$. Consistently, \autoref{fig:Pareto}(a) indicates that the chosen trade-off parameter $\gamma$ was effective: the overall optimum sits at the elbow of the front, delivering substantial drag reduction at an intermediate blowing level. A similar trend has been reported for the same van model when the actuation law is obtained by DRL. \citet{amico2024flow} trained agents either to maximize drag reduction or to maximize drag reduction under an energy penalty; in the former case, the solution used stronger bottom injection and produced the largest wake compression, whereas in the energy-penalized case, the actuation levels were lower and the wake remained closer to the baseline topology. The herein discussed optimization by HyGO reproduces this behavior (while obtaining better drag reduction, 8.8\% vs 5\% under actuation penalization) without changing the cost definition between experiments: early individuals sit near the “maximum-drag-reduction” part of the front (low $J_a$, high $J_b$), and subsequent generations move towards the energy-efficient branch while keeping $J_a$ nearly constant.

\autoref{tab:best_results} summarizes the metrics and settings for the overall best individual (minimum $J$), the reference cases (no actuation and steady-jet), and the individual that maximizes drag reduction (minimum $J_a$). The steady-jet reference is clearly inefficient with a cost value much higher than unity and a drag increase $J_a>1$. By contrast, the overall optimum achieves $J\approx1$, i.e., a net benefit even after penalizing mass injection. The configuration that maximizes drag reduction reaches $\approx 11\%$ reduction but does so at very high mass injection, resulting in $J>1$ and therefore worse than no actuation in terms of the mass injection–drag reduction balance. The global optimum strikes a better balance, delivering $\approx 8.8\%$ drag reduction while using roughly half of the available mass-flow budget, primarily by acting through the top and bottom slots.

\subsection{Base pressure analysis}
For a more detailed analysis of the flow dynamics, time-resolved pressure data were acquired for two-minute intervals at a sampling frequency of $f_s=800$~Hz. The resulting mean ($\overline{\Delta P} = \overline{P - P_\infty}$) and fluctuating ($\Delta P'=\overline{|\Delta P-\overline{\Delta P}|}$) differential pressure maps for these key scenarios are presented in \autoref{fig:Averages}. In the non-actuated case, the mean pressure map shows a pronounced low-pressure region near the lower part of the base. This feature is associated with the dominant shedding mechanism, which produces a large separated region, and the overall pressure distribution is consistent with previous studies \cite{amico2022geneticFurgo}. In contrast, all actuation strategies induce significant pressure recovery in this lower area, albeit at the expense of increased pressure losses near the upper edge. The best-performing (minimum $J$) and maximum drag reduction (minimum $J_a$) cases yield similar mean pressure profiles, characterized by a more extensive area of pressure recovery than that achieved with steady blowing, though with a lower peak pressure, supporting the hypothesis of the actuation targeting the lower separation region.

An analysis of the pressure fluctuations reveals that all control strategies increase the unsteadiness relative to the baseline case. Steady blowing, in particular, generates a large region of high-amplitude fluctuations in the central part of the base, a phenomenon likely attributable to the impingement and shedding of large-scale von K\'arm\'an-like vortices. Notably, the pulsed actuation cases produce even higher fluctuation levels, with an asymmetry biased towards the right side of the base. This spatial bias is hypothesized to be a direct consequence of the imposed phase shift between the lateral actuators, wherein the left-side jet systematically lags the right. Furthermore, the greater mass flow rate associated with the maximum drag reduction case (minimum $J_a$) correlates with a further increase in fluctuation intensity, including a distinct peak in $\Delta P'$ near the bottom of the measurement domain. In general, this spatial organization of the pressure fields is also consistent with the wake topologies reported by \citet{amico2024flow} for their DRL-controlled van. In their maximum-drag-reduction cases, the streamwise bubble is shortened, the wake becomes more symmetric, and a substantial pressure recovery is measured over the central and lower areas of the base. In our study, the minimal-$J$ and minimal-$J_a$ individuals generate the same signature (pressure gain low, mild losses high) but with an actuation found under an explicit mass-flow penalty. This supports our interpretation that HyGO is driving the flow toward the same physically efficient wake state identified by closed-loop DRL, only through an open-loop, model-free search.

Synchronous pressure and force measurements were acquired for all actuation cases (as detailed in~\autoref{s:IndividualEvaluation}). These measurements yielded mean $\overline{\Delta P}$ and fluctuating $\Delta P'$ pressure maps for each individual, averaged over the valid repetitions. To facilitate interpretation, the concatenated pressure data were processed using Multi-Dimensional Scaling (MDS)~\cite{cox2000mds, kaiser2017MDS}, a dimensionality reduction technique. MDS projects the high-dimensional data (i.e., 62 dimensions arising from 31 two-component pressure measurements) into a low-dimensional embedding. A correlation analysis between the control parameters and the embedding dimensions reveals that significant information is contained only within the first two dimensions, $\alpha_1$ and $\alpha_2$, allowing for a visual representation of the explored solution space during the optimization process. In the following, we stress that the embedding is constructed from pressure measurements, not from the actuation parameters, and therefore provides a flow-response-based representation of the optimization landscape.

\begin{figure*}
    \centering
    \includegraphics[width=1\linewidth]{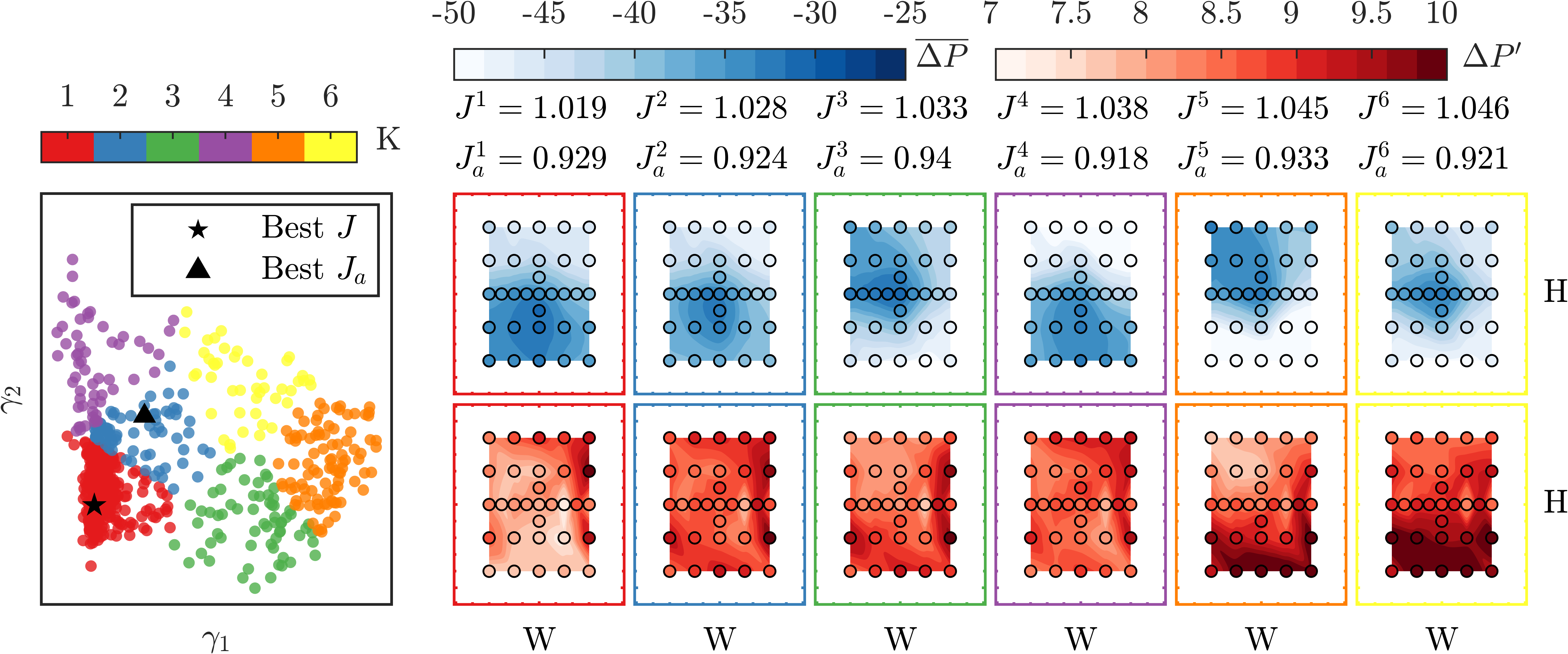}
    \caption{Identification of distinct control regimes via clustering. The left image displays the clustering of all evaluated control cases in the reduced-order space ($\alpha_1, \alpha_2$). The six identified groups are numbered and colored according to their mean cost $J$, from best-performing (Group 1) to worst-performing. The right panels include the weighted-average base pressure maps corresponding to each group. The top row displays the mean pressure ($\overline{\Delta P}$) and the bottom row shows the pressure fluctuations ($\Delta P'$). The maps are ordered by group number from left to right, revealing the characteristic flow topology associated with each performance level.}
    \label{fig:Clusters}
\end{figure*}

To interpret the low-dimensional embedding represented by $\alpha_1$ and $\alpha_2$, a clustering analysis is performed to distinguish different optimization regions. To ensure robustness, the regions are obtained by performing a K-neighbors clustering on a homogeneously downsampled embedding, which is then mapped to the full-dimensional data. The resulting six groups, illustrated in \autoref{fig:Clusters}, were ordered according to their mean cost function value, $\overline{J}^k$, computed as the average of the cost, $J_i$, for each individual $i$ in group $k$:
\begin{equation}\label{eq:cluster_weighting}
    \overline{J}^k = \frac{1}{N_k} \sum_{i=1}^{N_k} J_i
\end{equation}
where $N_{k}$ is the number of individuals in group $k$. We emphasize that the clustering is introduced primarily for visualization and interpretation: this low-order embedding represents a continuous family of pressure distributions, and the clusters (labelled groups to emphasize the method's target) serve as a compact discretization into representative regions rather than as strictly separated physical modes.

Visualizing the principal directions, $\alpha_1$ and $\alpha_2$, reveals significant trends in the optimization process (\autoref{fig:Clusters}). The high-performing individuals (i.e., those with lower cost function values) concentrate within a compact region of the low-dimensional embedding, providing a clear visualization of the optimization's convergence. As previously remarked, it is particularly noteworthy that this embedding was derived from the pressure measurements, not the actuation parameters themselves. This indicates a smooth relationship within the explored domain between the control inputs and the resulting pressure field, a feature effectively exploited by the algorithm to enhance convergence. Furthermore, the relative locations of the best $J$ and best $J_a$ solutions within the embedding indicate that the pressure manifold separates flow regimes not only according to aerodynamic performance, but also according to actuation-cost efficiency. The trade-off between $J_a$ and $J_b$ is also evident from the group-wise statistics: the fourth-best group according to $J$ is ranked third when evaluated in terms of $J_a$, indicating that this group achieves strong drag reduction at the expense of significantly higher mass injection. This embedding therefore reveals a preferential organization of solutions, with a predominantly vertical stratification associated with $J$, where lower regions correspond to actuation--cost performance, and a left--right organization correlated with $J_a$, reflecting better aerodynamic performance.

Following this classification, and averaging the pressure maps following the same procedure as in Equation~\eqref{eq:cluster_weighting}, the corresponding average mean and fluctuating pressure maps for each group are depicted in \autoref{fig:Clusters}. The results reveal a clear trend: higher drag-reduction groups are characterized by stronger mean pressure recovery concentrated over the lower portion of the base, together with comparatively reduced pressure fluctuations. Progressing from group 1 to group 6, the fluctuation levels $\Delta P'$ increase markedly, with the most intense fluctuations localized near the lower base region in the low-performance regimes. In parallel, the mean-pressure footprint becomes less favorable on the lower base and appears increasingly localized and/or shifted upward in the worst-performing groups.
An exception to this trend is observed in the fourth group, which exhibits a $\overline{\Delta P}$ distribution similar to that of the best-performing group due to its large drag reduction (low $J_a$); however, the higher mass injection leads to increased pressure fluctuations (as evidenced in the $\Delta P'$ map), placing this group in the fourth position in terms of $J$.
Overall, these group-wise pressure signatures provide a compact overview of the wake regimes explored by the optimizer and their association with aerodynamic performance.

\begin{figure*}
    \centering
    \includegraphics[width=0.95\linewidth]{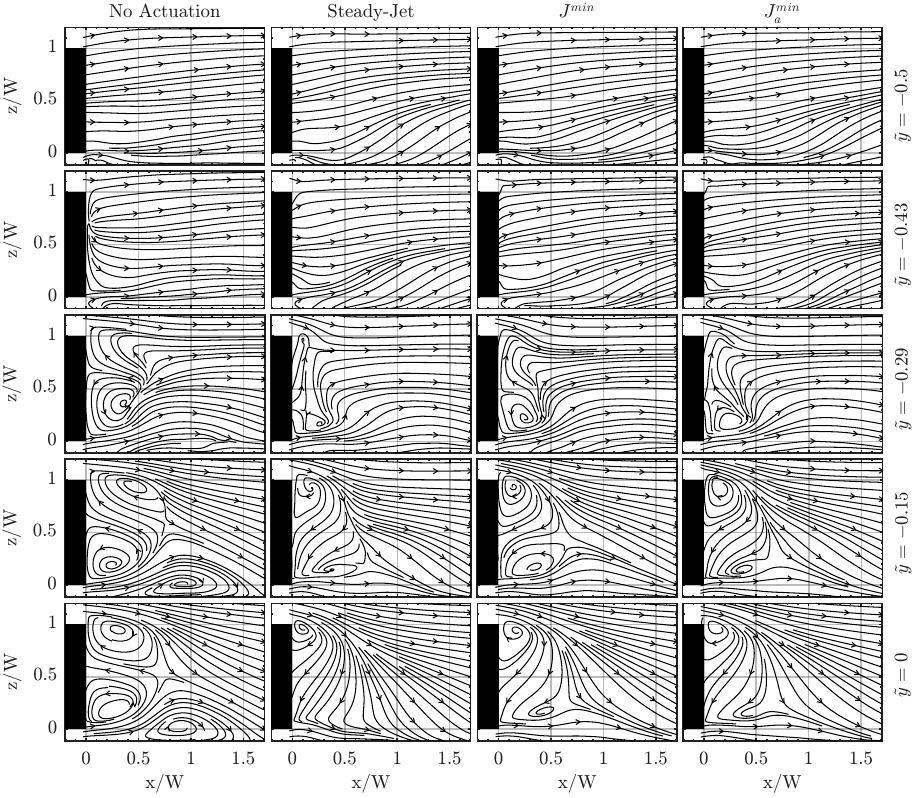}
    \caption[]{Time-averaged streamlines for the four key control scenarios. Each row corresponds to a different scenario as detailed in \autoref{tab:best_results}: (from left to right) no actuation, steady-jet, best-performing $J$, and best-performing $J_a$. Each column represents a different vertical measurement plane at the spanwise locations of (from right to left) $y/W = 0, -0.15, -0.29, -0.43,$ and $-0.5$.}
    \label{fig:PIV_streamlines}
\end{figure*}
\subsection{Flow field analysis}\label{s:piv_analysis}
\begin{figure*}
    \centering
    \includegraphics[width=0.95\linewidth]{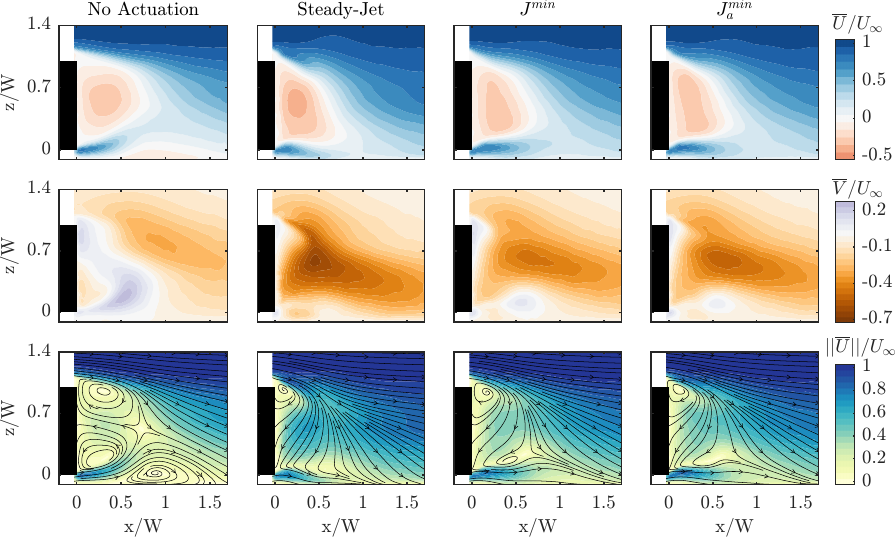}
    \caption{Time-averaged velocity fields in the symmetry plane ($y/W=0$) for the four key control scenarios detailed in \autoref{tab:best_results}. The columns correspond to each scenario. The rows display contours of: \textbf{(top)} mean streamwise velocity, $\overline{U}/U_\infty$; \textbf{(middle)} mean vertical velocity, $\overline{V}/U_\infty$; and \textbf{(bottom)} velocity magnitude, $||\overline{U}||/U_\infty$, with superimposed streamlines.}
    \label{fig:PIV_means_y0}
\end{figure*}
To investigate the physical mechanisms of control, a PIV analysis was performed for the four key scenarios identified in \autoref{tab:best_results}. For each case, 500 velocity fields were measured in nine vertical planes, as depicted in \autoref{fig:exp_setup_piv}. The velocity fields are normalized by the free-stream velocity, $U_{\infty}$. The mean streamlines for the five central planes of the uncontrolled baseline case are presented in \autoref{fig:PIV_streamlines}; due to the flow's symmetry, only one side is shown. The baseline wake structure is consistent with previous studies on similar bluff bodies \cite{grandemange2013turbulent_wake_modes_bistability_flat_back, fan2020experimentalBiStableFlatBackAhmed, dalla2019simulationsBimodalFlatBackBluff,amico2024flow,amico2022deep,amico2022geneticFurgo}, featuring a saddle point at $z/W\approx 0.5$ that separates two primary counter-rotating vortices within the main recirculation region.
A smaller, wall-proximate vortex structure is also observed beneath (and downstream) the main bubble, consistent with the upward velocity induced by the lower half of the primary recirculation.
The three-dimensional extent of this recirculation bubble is confined to the central portion of the base ($y/W\lesssim 0.4$), as it is no longer present in the planes aligned with the actuators ($y/W=0.43$). While the bubble's length remains constant between the centerline and the first off-center plane ($y/W=0.15$), the core of the top vortex is displaced forward. Further outboard ($y/W=0.29$), this top vortex disappears, accompanied by a rapid decrease in the bubble's overall length. In the outermost planes ($y/W \geq 0.43$), the flow is aligned with the freestream, though a complete analysis of this region, likely dominated by longitudinal trailing vortices \citep{liu2021flowAhmed3D}, is precluded by the two-dimensional nature of the measurements.

The effect of actuation is clearly evident in the three central PIV planes ($|y|/W=[0,0.15,0.29]$) shown in \autoref{fig:PIV_streamlines}. All three control strategies introduce a significant negative vertical velocity ($w$) into the near-wake, which profoundly alters the baseline topology. This downwash flow deflection towards the wall is highest at the steady-jet case, followed by the minimal $J_a$, suggesting a strong dependence on the injected mass flow. Such momentum injection significantly dampens the bottom recirculation vortex, eliminating it entirely at the centerline in the steady-jet case, while concurrently reducing the size of the top vortex. Furthermore, both the best-performing $J$ and $J_a$ cases displace the weakened bottom vortex away from the model's base, which is consistent with the increased base pressure suggested by the measurements in \autoref{fig:Averages}. This localized pressure growth is maximum at the central plane, which seems to be the most affected by the actuations, following the previous flow description. This pattern of bubble compression and vertical reorganization of the recirculation is similar to that observed for the DRL-selected forcings on the identical van model \cite{amico2024flow} (in the penalized scenario). In their Cases 1–2, the DRL agent injected momentum from the lower edge so as to “anticipate” the interaction of the upper and lower shear layers, thereby moving the saddle point upstream and closer to the base and yielding a symmetric wake and high base pressure. Our best-performing-$J$ actuates similarly: the center-plane PIV shows (i) suppression of the bottom vortex, (ii) contraction of the main bubble, and (iii) strengthening of the downward motion impinging on the base (\autoref{fig:PIV_streamlines}), all of which explain the pressure recovery in \autoref{fig:Averages}. Similarly, it is highest for the steady-jet case, a condition consistent with the observed flow and pressure recovery in \autoref{fig:Averages}. Similarly, the top vortex is brought closer to the model's base in all three actuated cases, suggesting its involvement in the pressure drop observed previously in this region (\autoref{fig:Averages}).
\begin{figure*}
    \centering
    \includegraphics[width=0.95\linewidth]{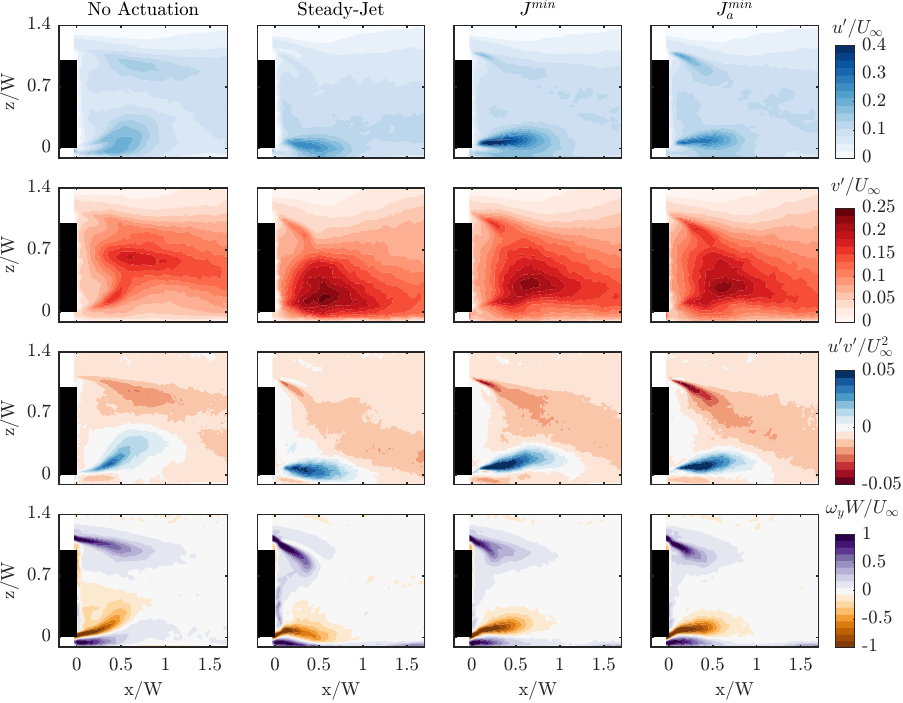}
    \caption{Second-order turbulence statistics in the symmetry plane ($y/W=0$) for the four key control scenarios detailed in \autoref{tab:best_results}. The columns correspond to each scenario. The rows, from top to bottom, display contours of: \textbf{(first)} streamwise velocity fluctuations, $u'/U_\infty$; \textbf{(second)} vertical velocity fluctuations, $v'/U_\infty$; \textbf{(third)} Reynolds shear stress, $u'v'/U_\infty^2$; and \textbf{(fourth)} mean spanwise vorticity, $\omega_y\cdot W/U_\infty$.}
    \label{fig:PIV_second_order_y0}
\end{figure*}

This modification of the primary vortical structures leads to a substantial reorganization of the entire recirculation region. The saddle point is displaced downwards and forwards, reducing both the length and width of the recirculation bubble, with the latter being most apparent in the $y/W=0.29$ plane. This downward shift of the stagnation point also contributes to the elimination of the small wall-attached vortex, promoting boundary layer re-attachment and likely driving the observed drag reduction. Finally, in the outboard planes ($y/W \geq 0.43$), the actuated cases exhibit an upward deflection of the flow downstream of $x/W > 0.5$. This suggests the formation of more pronounced longitudinal trailing vortices, a hypothesis that would require further investigation with horizontal PIV planes to confirm.

Further analysis of the flow statistics in the symmetry plane ($y/W=0$) provides deeper physical insight into the control mechanisms. The time-averaged streamwise velocity contours ($\overline{U}$), presented in \autoref{fig:PIV_means_y0}, clearly depict the actuation-induced modifications to the recirculation bubble. These fields highlight the effect of the bottom jet, which significantly alters the departure angle of the flow from beneath the bluff body; stronger actuation correlates with a more pronounced upward deflection.

As previously discussed, all control cases reduce the length of the recirculation bubble while simultaneously increasing the magnitude of the reversed flow velocity within it. A similar trend is observed in the vertical velocity ($\overline{V}$) contours, which show a substantial increase in downward velocity that suppresses the wall-bounded vortex. The magnitude of this effect is greatest for the steady-jet case, followed by the minimum $J_a$ case, indicating a strong correlation with the injected momentum. Furthermore, the $\overline{V}$ fields reveal a small region of positive velocity near the wall, coinciding with the location of the bottom recirculation vortex. The intensity of this upward velocity is a direct indicator of the vortex's circulation, confirming its suppression in the controlled cases.

An analysis of the second-order statistics, presented in \autoref{fig:PIV_second_order_y0}, provides further insight into the control mechanisms. The contours of streamwise velocity fluctuations ($u'$) confirm that the most significant turbulent activity originates from the bottom jet in all actuated scenarios. Interestingly, the highest intensity of streamwise fluctuations is generated by the best-performing control case (minimum $J$). In contrast, the case with the largest drag reduction (minimum $J_a$) trades these bottom fluctuations for a more even distribution between the bottom and bottom slots, as evidenced by the significantly higher Reynolds stresses ($u'v'$).

Linking the control analysis with the associated flow-field modifications observed for the best-performing actuation, the bottom slot operates at a low frequency ($\tilde{f}_3 \approx 0.12$), close to the natural vortex-shedding mode of the wake ($\tilde{f}_{\mathrm{sh}} \approx 0.17$ in the present configuration). Within the classical framework of bluff-body flow control, actuation in this frequency range is typically associated with interaction with the global wake instability, affecting vortex formation and base pressure \cite{choi2008control}. Consistently, the time-averaged PIV fields reveal wake narrowing, recirculation shortening, and base-pressure recovery, which are canonical signatures of large-scale wake modification. The top slot, on the other hand, operates at a much higher frequency ($\tilde{f}_1/\tilde{f}_3 \approx 7.08$), well separated from the shedding scale. According to established separation-control literature, forcing in this regime is generally associated with the response of unstable shear layers rather than with direct excitation of the global mode \cite{GREENBLATT2000lowfreq,choi2008control}. The proximity of the top slot to the separation region and the energization of the Reynolds stresses and vorticity in the top edge of the planar fields further supports the interpretation that this actuation may influence the early shear-layer development and, indirectly, the downstream wake organization. It is emphasized, however, that in the absence of time-resolved velocity or pressure spectra, and the planar nature of the presented PIV measurements, these mechanisms cannot be isolated experimentally in the present work. The proposed interpretation should therefore be regarded as a physically consistent explanation grounded in established flow-control theory, rather than as a direct demonstration of targeted modal interactions.

\section{Conclusions}\label{s:conclusions}
This work demonstrates an experiment-in-the-loop, model-free optimization of open-loop pulsed-jet actuation for reducing the aerodynamic base drag of a generic road-vehicle model. The optimization was carried out using a Hybrid Genetic Algorithm (HyGO), which effectively navigated a large parameter space to identify a local-optimal, non-intuitive control law.

The primary finding of this study is a significant and robust aerodynamic drag reduction of approximately 8.8\%. This control strategy was identified using a cost function carefully designed to ensure a net energy saving by simultaneously targeting drag minimization and penalizing the energy expenditure of the actuation system. The reliability of this solution was confirmed through repeated tests, which consistently reproduced the drag reduction. Crucial physical insight into the control mechanism was provided by synchronous base pressure measurements. These revealed that the actuation concentrates its effect on the lower half of the model's base, leading to a substantial pressure recovery in this region, which is the main driver for the overall drag reduction.

A detailed analysis of the best-performing control law reveals a clear differentiation in the role of the actuators. The main contributor to the drag reduction is the bottom jet, which operates at low frequencies to counteract the primary vortex shedding mechanism responsible for the bulk of the base pressure deficit. In contrast, the top jet focuses on higher-frequency, less energetic phenomena, suggesting a role in disrupting smaller-scale turbulent structures in the upper shear layer. It is also important to reflect on the architecture of the chosen cost function. The formulation of the term penalizing the injected mass flow ($J_b$) implies that a single duty cycle parameter for the lateral jets affects the cost function twice as much as those for the top and bottom jets. This structure may have artificially dampened the actuation from the side jets, pushing the optimizer towards solutions that minimized their contribution. Future studies could consider a modified cost function that decouples the penalty for each side actuator, which might unlock different and potentially more effective control strategies involving lateral forcing.

Finally, the most efficient solutions discovered here reproduce the same wake archetype (shorter, high-pressure base) that appears when the same van model is controlled by DRL agents trained with and without an energy term \cite{amico2024flow}. This consistency between HyGO, as an open-loop optimizer, and a DRL agent, as a closed-loop controller, AI-based controller strengthens the generality of the proposed mechanism.

\section*{Acknowledgments}
The authors gratefully acknowledge Professors A. Ianiro and S. Discetti (UC3M) for their insightful comments and suggestions. The contributions of E. Amico, Professors J. Serpieri and G. Cafiero (PoliTO), and Jose Luis Oterino (INTA) in assisting with the experimental setup and van model, as well as the technical support of D. Bergmann with the wind-tunnel operation, are also sincerely appreciated.

\section*{Declaration of generative AI and AI-assisted technologies in the manuscript preparation process}

During the preparation of this work, the authors used ChatGPT, Gemini and Grammarly in the writing process to improve the readability and language of the manuscript. After using this tool, the authors reviewed and edited the content as needed and assume full responsibility for the content of the published article.

\section*{Data availability}
Data will be made available upon request.

\section*{Funding sources}
This activity was funded by project ACCREDITATION (Grant No TED2021-131453B-I00), funded by MCIN/AEI/ 10.13039/501100011033 and by the ``European Union NextGenerationEU/PRTR''.


\bibliographystyle{arxiv_bib}
\bibliography{bib}

\newpage
\appendix
\section{Individual evaluation process}\label{s:IndividualEvaluation}
This subsection details the experimental procedure for evaluating each individual. The primary objective is to accurately determine the aerodynamic cost, $J_a$, and the actuation penalty, $J_b$, for each control strategy while mitigating experimental uncertainties. A significant challenge is potential drift in the load-cell readings, as the experiments lasted up to 10 hours on the same day. Although this drift remains small ( below $0.1\%$ F.S. per $10^\circ$) thanks to the use of a temperature-compensated load cell and a climate-controlled environment, it can still affect the optimization process by introducing an additional source of uncertainty. To counteract this effect and ensure robust measurements, a self-referencing procedure was implemented. For each actuated case, a corresponding baseline (no-actuation) drag measurement, $D_0$, was taken immediately before the actuated drag measurement, $D$. This approach ensures that the resulting aerodynamic cost, calculated as the ratio $J_a = D/D_0$, is insensitive to potential low-frequency drifts, thereby stabilizing the evaluation against time-of-day variations.

Other uncertainties (e.g., sensor noise and transient responses) are reduced by introducing a stabilization (wait) time $T_w$ after any change in actuation, and by using a measurement window $T_m$ long enough to ensure statistical convergence. Even with these adjustments, ratio-type metrics such as $J_a$ can produce occasional outliers that could mislead HyGO if they appear among the best candidates.

To guard against this, each individual is measured twice by default as follows. First, we acquire the no-actuation baseline after a wait $T_{w,D_0}=12~\mathrm{s}$, then record for $T_{m,D_0}=20~\mathrm{s}$, logging drag and base pressure synchronously. Next, we apply the individual’s actuation, wait $T_{w,D}=12~\mathrm{s}$, and record for $T_{m,D}=20~\mathrm{s}$, logging drag, base pressure, and mass-flow rate. 

From each repetition $r$, we compute $J_a^{r} = D^{r}/D_0^{r}$. If the relative discrepancy $\sigma_{12} \;=\; \big|J_a^{1}-J_a^{2}\big|/\max\big(J_a^{1},J_a^{2}\big)$) exceeds  $2.5\%$ , a third repetition is performed. We then compute the three pairwise discrepancies 
$\sigma_{12},\sigma_{13},\sigma_{23}$ and retain the pair with the smallest 
$\sigma$; the corresponding $J_a$ (and $J_b$, computed from the associated actuated segments) are averaged and used to form the final cost $J$. If, despite three repetitions, the minimum pairwise discrepancy remains above $2.5\%$, the individual is discarded by assigning an extreme cost $J=10^{36}$. The full evaluation routine is summarized in Algorithm~\ref{alg:evaluate_cost}.
\begin{algorithm}
    \caption{Cost Function Evaluation}
    \label{alg:evaluate_cost}
    \begin{algorithmic}[1]
        \REQUIRE Parameter vector $\theta=[\tilde f_1,DC_1,\tilde f_2,DC_2,\tilde f_3,DC_3,\phi]$
        \item[]
        \FOR{$r=1$ \TO $2$}
            \item[] \textit{--- Measure No Actuation ---}
            \STATE Turn off valves
          \STATE Wait $T_{w,D_0}$ for stabilization
          \STATE Measure $D_0$ and $P_0$ for $T_{m,D_0}$ seconds
          \item[] \textit{--- Measure Actuation ---}
          \STATE Set valves to $\theta$
          \STATE Wait $T_{w,D}$ for stabilization
          \STATE Measure $D$, $\dot m$, and $P$ for $T_{m,D}$ seconds
          \item[] \textit{--- Compute Costs ---}
          \STATE $J_a^r = \overline{D}/\overline{D_0},\;J_b^r=\overline{\dot m}/\dot m_{SJ}$
        \ENDFOR
        \STATE Compute uncertainty $\sigma_{12} = \frac{|Ja^1 - J_a^2|}{max(Ja^1,J_a^2)}$.
        \item[]
        \IF{$\sigma_{12}<0.025$}
            \STATE $J = \frac{Ja^1 + J_a^2}{2} + \gamma \frac{Jb^1 + J_b^2}{2}$
        \ELSE
            \item[] \textit{--- Measure No Actuation ---}
            \STATE Turn off valves
          \STATE Wait $T_{w,D_0}$ for stabilization
          \STATE Measure $D_0$ and $P_0$ for $T_{m,D_0}$ seconds
          \item[] \textit{--- Measure Actuation ---}
          \STATE Set valves to $\theta$
          \STATE Wait $T_{w,D}$ for stabilization
          \STATE Measure $D$, $\dot m$, and $P$ for $T_{m,D}$ seconds
          \item[] \textit{--- Compute Costs ---}
          \STATE $J_a^3 = \overline{D}/\overline{D_0},\;J_b^3=\overline{\dot m}/\dot m_{SJ}$
          \STATE Compute uncertainty $\sigma_{13} = \frac{|Ja^1 - J_a^3|}{max(Ja^1,J_a^3)},\; \sigma_{23} = \frac{|Ja^2 - J_a^3|}{max(Ja^2,J_a^3)}$.
          \item[]
            \IF{$\sigma_{13}\leq\sigma_{23}$ \textbf{AND} $\sigma_{13}<0.025$}
                \STATE $J = \frac{Ja^1 + J_a^3}{2} + \gamma \frac{Jb^1 + J_b^3}{2}$
            \ELSIF{$\sigma_{23}<\sigma_{13}$ \textbf{AND} $\sigma_{23}<0.025$}
                \STATE $J = \frac{Ja^2 + J_a^3}{2} + \gamma \frac{Jb^2 + J_b^3}{2}$
            \ELSE
                \STATE $J=10^{36}$
            \ENDIF
        \ENDIF
    \end{algorithmic}
\end{algorithm}

The $2.5\%$ uncertainty threshold was established from a dedicated preliminary repeatability campaign designed to characterize the variability of the measurement process. Three representative individuals, spanning distinct regions of the actuation space, were each evaluated ten times. Each evaluation included both baseline ($D_0$) and actuated segments, ensuring like-for-like force comparisons consistent with the optimization protocol. Rather than relying solely on the limited number of individuals, the statistical population was constructed from the combinatorial set of pairwise discrepancies between repeated measurements. For each individual, the ten repetitions yield $45$ independent pairwise comparisons (each measurement is compared with the other 9), resulting in a total population of $135$ discrepancy values. The distribution of these discrepancies (or uncertainties) was analyzed, and the $95^{\mathrm{th}}$ percentile was found to lie at approximately $2.45\%$. This value was rounded to $2.5\%$ and adopted as the operational threshold. In statistical terms, this implies that approximately $5\%$ of evaluations are expected to exceed the threshold due to stochastic experimental variability alone, thereby triggering a third repetition rather than immediate rejection. In fact, $7.6\%$ of the evaluations required a third repetition, in close agreement with the expected confidence interval given the broader range of flow conditions explored, and notably, no individuals were ultimately discarded. This confirms that the selected threshold was sufficient to filter unreliable measurements while avoiding the elimination of genuinely high-performing solutions.

The selected stabilization and acquisition windows, together with processing and communication latencies, yield an average evaluation time of $\approx145~\mathrm{s}$ and $\approx220~\mathrm{s}$ per individual for two and three repetitions respectively. The repetition protocol brought all evaluated individuals within the $2.5\%$ uncertainty criterion, with a mean minimum pairwise discrepancy of $1.03\%$. Of the 688 individuals assessed, 52 ($\approx7.6\%$) required a third repetition; none were discarded afterwards.

\end{document}

%% file: utils/commands.tex
\makenomenclature
\RestyleAlgo{ruled}

\definecolor{blue}{rgb}{0,0,1}
\definecolor{red}{rgb}{1,0,0}
\definecolor{black}{rgb}{0,0,0}
\definecolor{white}{rgb}{1,1,1}
\definecolor{grey}{RGB}{50,50,50}
\definecolor{redR}{RGB}{227, 47.3333, 39}
\definecolor{green}{RGB}{55, 160.3333, 85}
\definecolor{blue}{RGB}{55, 135, 192.3333}
\definecolor{yellow}{rgb}{0.9412, 0.7843, 0.0588}

\definecolor{static}{rgb}{0.0 0.8 0.4}
\definecolor{rocket}{rgb}{0.9 0.3 0.3}
\definecolor{satellite}{rgb}{0.2 0.6 1.0}
\definecolor{mutation}{RGB}{204, 153, 255}
\definecolor{crossover}{RGB}{0, 204, 153}

\colorlet{static_violin}{static!30!white}
\colorlet{rocket_violin}{rocket!30!white}
\colorlet{satellite_violin}{satellite!30!white}

\definecolor{color_rebut}{RGB}{20, 20, 255}

\newcommand{\HYGO}{\textsf{HyGO}\xspace}
\definecolor{yellow_star}{rgb}{0.9961, 0.6774, 0.2167}
\definecolor{myblue}{rgb}{0., 0.4, 1}
\def\tsc#1{\csdef{#1}{\textsc{\lowercase{#1}}\xspace}}
\tsc{WGM}
\tsc{QE}